\documentclass[a4paper, 11pt]{article}
\usepackage{comment} 
\usepackage[nointegrals]{wasysym}
\usepackage{fullpage} 
\usepackage{graphicx}
\usepackage{cite}
\usepackage{bm}
\usepackage{float}
\usepackage{amsmath}
\usepackage[utf8]{inputenc}
\usepackage{amssymb}
\usepackage{mathrsfs}
\usepackage{wrapfig}
\usepackage{enumitem}
\usepackage[english]{babel}
\usepackage{comment}
\usepackage{bbold}  
\usepackage{multicol}
\usepackage[usenames,dvipsnames]{xcolor} 
\usepackage{xcolor,hyperref}
\usepackage{setspace}
\usepackage{abstract}
\usepackage{sectsty}
\sectionfont{\large}
\usepackage{tablefootnote}

\usepackage[T1]{fontenc}
\usepackage[margin=0.94in]{geometry}
\begin{document}

\null\hfill 
KEK-TH-2823, KEK-Cosmo-0416 \\

\vspace*{\fill}
\begin{center}
    \Large\textbf{{ \textcolor{Black}{Probing Majoron Dark Matter with \\ Gravitational Wave Detectors} }}\\
    \vspace{1cm}
    \normalsize\textsc{Ippei Obata}$^{a,b}$ and \textsc{Tsutomu T. Yanagida}$^{b,c}$ 
\end{center}

\begin{center}
$^{a}$ \textit{Theory Center, Institute of Particle and Nuclear Studies (IPNS), High Energy Accelerator Research Organization (KEK), 1-1 Oho, Tsukuba, Ibaraki 305-0801, Japan}\\
$^{b}$ \textit{Kavli IPMU (WPI), UTIAS, The University of Tokyo, Kashiwa 277-8583, Japan}\\
$^{c}$ \textit{Tsung-Dao Lee Institute \& School of Physics and Astronomy, Shanghai Jiao Tong University, Pudong New Area, Shanghai 201210, China}
\end{center}
\thispagestyle{empty} 

\renewcommand{\abstractname}{\textsc{\textcolor{Black}{Abstract}}}

\begin{abstract}
The Majoron is a hypothetical (pseudo) Nambu-Goldstone boson arising from the spontaneous breaking of a global lepton number symmetry, and is known as a candidate for dark matter in our Universe. In this paper, we investigate the possibility of probing the Majoron dark matter with a linear optical cavity used in the interferometric gravitational wave detectors. We consider a scenario in which the Majoron dark matter couples to photons through a QED anomaly, leading to an oscillatory photon birefringence induced by the coherent dark matter background. The anomaly coefficient is fixed by requiring the model to simultaneously reproduce the electroweak Higgs scale and a typical right-handed Majorana neutrino mass scale, and the resulting dark matter-photon coupling naturally falls within the sensitivity range of optical interferometers. By incorporating additional optics to extract the birefringence signal, we find that ground-based laser interferometers such as Advanced LIGO, KAGRA, as well as future detectors, can probe a region of the parameter space of Majoron dark matter.
\end{abstract}
 
\vfill

\small
\href{mailto:iobata@post.kek.jp}{iobata@post.kek.jp}\\
\href{mailto:tsutomu.tyanagida@sjtu.edu.cn}{{tsutomu.tyanagida@sjtu.edu.cn}}
\vspace*{\fill}

\clearpage
\pagenumbering{arabic} 
\newpage

\tableofcontents

\section{Introduction}

The existence of small neutrino masses requires physics beyond the Standard Model. One of the most compelling scenarios is the seesaw mechanism \cite{Minkowski:1977sc,Yanagida:1979as,Yanagida:1979gs,Glashow:1979nm,Gell-Mann:1979vob},\footnote{The term \textit{seesaw mechanism} was originally introduced by one of the authors at the INS symposium in Tokyo in 1981~\cite{INS1981}, where the dimension-five operator for neutrino masses was also indicated.} which elegantly explains the smallness of neutrino masses by introducing heavy Majorana right-handed neutrinos and their mass mixing with the Standard Model neutrinos. Since Majorana neutrinos violate lepton number, this framework also provides a natural mechanism for generating the baryon asymmetry of the Universe via leptogenesis \cite{Fukugita:1986hr,Buchmuller:2005eh}.
Despite its success, the origin of the heavy Majorana masses remains unclear. A well-motivated possibility is that they arise from spontaneous symmetry breaking at a high energy scale. If the underlying symmetry is gauged, one typically obtains a massive gauge boson associated with an anomaly-free baryon-minus-lepton ($B-L$) symmetry \cite{Wilczek:1979hh,Barbieri:1979ag}. On the other hand, if the symmetry is global, its spontaneous breaking leads to a Nambu-Goldstone boson carrying lepton number, known as the Majoron \cite{Chikashige:1980qk,Chikashige:1980ui,Gelmini:1980re}, in addition to light neutrinos.
An intriguing possibility is that the Majoron acquires a small mass through explicit breaking of the global symmetry, for instance by soft terms or quantum gravitational effects \cite{Giddings:1988cx,Coleman:1988tj}.
In such a case, the Majoron can serve as a viable dark matter candidate \cite{Akhmedov:1992hi,Rothstein:1992rh,Berezinsky:1993fm,Gu:2010ys,Frigerio:2011in}.

Conventionally, the Majoron has been thought to interact predominantly with neutrinos.
In the original model, heavy Majorana neutrinos couple to the lepton doublets and the Higgs doublet of the Standard Model, and the Majoron arises as a Nambu-Goldstone boson associated with a global lepton number symmetry.
Due to the charge assignment of the global lepton number, the Majoron does not possess a QED anomaly and hence does not couple to photons.
Recently, however, a new scenario in which the Majoron possesses a QED anomaly has been proposed \cite{Liang:2024vnd,Lu:2025kbe}.
In this scenario, an additional Higgs doublet with a nontrivial global $U(1)$ charge, which couples to lepton doublets and heavy Majorana neutrinos, is introduced.
Then, as in QCD axion models \cite{Dine:1981rt,Zhitnitsky:1980tq,Langacker:1986rj}, the presence of two Higgs doublets enables leptons to carry opposite charges under the global symmetry, thereby inducing a QED anomaly.
Owing to this anomaly, the Majoron couples to photons via a topological Chern-Simons interaction, making it potentially testable in a variety of axion experiments.
Interestingly, assuming that the Majoron constitutes the entirety of dark matter, the conventional thermal leptogenesis scenario predicts a dark matter mass below the $\mu\mathrm{eV}$ scale \cite{Buchmuller:2005eh}.
This parameter region, characterized by the Majoron mass and its photon coupling, overlaps with that of QCD axion dark matter and can be probed by axion haloscope experiments such as ADMX \cite{Sikivie:1983ip,ADMX:2009iij,ADMX:2018gho,ADMX:2019uok,ADMX:2021nhd}.
Furthermore, unlike the QCD axion, the Majoron dark matter-photon coupling exhibits only a mild dependence on its mass \cite{Liang:2024vnd}.
Therefore, not only in the $\mu\mathrm{eV}$ range but also at lower masses, there remains significant potential to probe Majoron dark matter using a broader range of experimental approaches.

Inspired by the above works, we revisit the viable parameter space of anomalous Majoron models and explore new possibilities for probing Majoron dark matter using broadband detection methods.
As in the case of axions, the Majoron dark matter-photon interaction induces a difference in the phase velocities of circularly polarized photons, leading to an oscillatory rotation of the polarization angle with a frequency set by the dark matter mass.
This photon birefringence effect can be probed by a variety of optical ring cavity experiments \cite{Obata:2018vvr,Michimura:2019qxr,Oshima:2021irp,Fujimoto:2021uod,Oshima:2023csb,Takidera:2025hcw,Liu:2018icu,Pandey:2024dcd,Martynov:2019azm,Heinze:2023nfb,Heinze:2024bdc}.
In addition, laser interferometers developed for gravitational wave detection can also serve as powerful probes of dark matter owing to their large-scale infrastructure and advanced optical technologies.
With appropriate modifications to extract polarization birefringence signals from the resonant cavities, these detectors can be utilized for dark matter searches \cite{DeRocco:2018jwe,Nagano:2019rbw,Nagano:2021kwx,Nakatsuka:2022gaf,Gue:2024txz,Morisaki:2025bjs}\footnote{In particular, the KAGRA interferometer has already implemented dedicated optics for dark matter searches \cite{Michimura:2021hwr,Michimura:2025kod}.
}.
While these tabletop experiments provide excellent sensitivity at lower frequencies, gravitational wave detectors can potentially achieve superior sensitivities at higher frequencies, making them particularly well suited for Majoron dark matter.

This paper is organized as follows.
In Section~\ref{section: model setup}, we revisit the anomalous Majoron model and derive the corresponding photon coupling constant.
In Section~\ref{section: cosmo}, we describe the cosmological evolution of the Majoron field and evaluate the relic dark matter abundance, establishing the relation between the coupling constant and the dark matter mass.
In Section~\ref{section: main result}, we present the parameter space of Majoron dark matter that can be probed by future gravitational wave detectors.
We conclude in Section~\ref{section: conclusion}, where we also discuss future prospects.
Throughout this paper, we adopt natural units with $\hbar = c = 1$.

\section{Model setup} \label{section: model setup}

In this section, we present an anomalous Majoron model and evaluate the electromagnetic coupling of the Majoron. 
The relevant Lagrangian in the present model includes the following terms
\begin{align}
\mathcal{L} &\supset \bar{q}_L Y_u u_R \tilde{H}_1 + \bar{q}_L Y_d d_R H_1 + \bar{\ell}_LY_e e_R H_2 + \bar{\ell}_LY_D N_R \tilde{H}_1  \notag \\
&+ \dfrac{1}{2}\bar{N}^c_RY_NN_R\Phi^* + c_\Phi \dfrac{\Phi^n}{M_p^{n-2}}H_2^\dagger H_1 + V(H_1, \ H_2, \ \Phi) + \rm h.c. \label{eq: model}
\end{align}
consist of Yukawa interactions, Majorana mass term, and an effective higher-dimensional operator.
The model assumes a global $U(1)_N$ symmetry where the right-handed neutrinos $N_R$ have a $U(1)_N$ charge $1/2$. 
And we introduce a scalar boson $\Phi$ whose vev gives large Majorana masses for the right-handed neutrinos.
The phase of $\Phi \propto e^{iJ(x)/F_J}$ is nothing but the Majoron $J(x)$.
The higher dimensional operator $\Phi^n H_2^\dagger H_1/M_p^{n-2}$ is the key point in our extension, while the original model considers $n=1$ \cite{Liang:2024vnd}.
This operator contributes to an off-diagonal term of the Higgs mass matrix, whose mixing mass $\delta^2$ is evaluated as
\begin{equation}
\delta^2 \sim \dfrac{F_J^{n}}{M_p^{(n-2)}} \ .
\end{equation}
Assuming $c_\Phi = \mathcal{O}(1)$ and a typical symmetry breaking scale $F_J = \mathcal{O}(10^{14}) \ \text{GeV}$, $\delta$ becomes electroweak scale $\mathcal{O}(10^2-10^3) \ \text{GeV}$ at $n = 8$.

In Table \ref{tab:contents}, we list up the contents of quarks, leptons, Higgs, and the complex scalar particle and their charges associated the gauge group of Standard Model and $U(1)_N$.
\begin{table*}[htb]
\centering
 \caption{Field contents and $U(1)_N$ charge.
 }
 \label{tab:contents}
 \begin{tabular}{c|c|c|c|c}
  \text{Field} & $SU(3)_c$ & $SU(2)_L$ & $U(1)_Y$ & $U(1)_N$  
  \\ \hline
     $q_L$ & 3 & 2 & 1/6 & $\chi_q$ \\ \hline
     $u_R$ & 3 & 1 & 2/3 & $\chi_q + \chi_1$ \\ \hline
     $d_R$ & 3 & 1 &-1/3 & $\chi_q - \chi_1$ \\ \hline
  $\ell_L$ & 1 & 2 &-1/2 & 1/2 - $\chi_1$ \\ \hline
     $e_R$ & 1 & 1 &  -1 & 1/2 - $n$ - 2$\chi_1$ \\ \hline
     $N_R$ & 1 & 1 &   0 & 1/2 \\ \hline
     $H_1$ & 1 & 2 &   1/2 & $\chi_{1}$ \\ \hline
     $H_2$ & 1 & 2 &   1/2 & $\chi_{1} + n$ \\ \hline
    $\Phi$ & 1 & 1 &   0 & 1
 \end{tabular}
\end{table*}
With these charge assignments, this model produces QED anomaly but not QCD anomaly:
\begin{equation}
\mathcal{L} \supset \dfrac{1}{4}g_{J\gamma} J F\tilde{F} \ . \label{eq: CS}
\end{equation}
The expression for $g_{J\gamma}$ is given by:
\begin{equation}
g_{J\gamma} = c_{J\gamma}\dfrac{\alpha}{\pi}\dfrac{1}{F_J} \ ,
\end{equation}
where $c_{J\gamma}$ is an anomaly constant and $\alpha = 1/137$ is the QED fine structure constant.
Then, the anomaly coefficient is evaluated as $c_{J\gamma} = 3n$ \cite{Liang:2024vnd}.

\section{Cosmological evolution of Majoron background} \label{section: cosmo}

In this section, we estimate a relic abundance of Majoron dark matter by assuming a misalignment production mechansim \cite{Arias:2012az}, similar with that of axion-like particle.
Regarding the potential of Majoron, we consider a periodic form:
\begin{equation}
V(J) = m_J^2F_J^2\left[ 1 - \cos\left(\dfrac{J}{F_J}\right) \right] \ ,
\end{equation}
where $m_J$ and $F_J$ are the mass and decay constant of Majoron field $J$.
In Friedmann-Lema\^\i tre-Robertson-Walker metric spacetime $ds^2 = -dt^2 + a(t)^2d\bm{x}^2$, the equation of motion for the Majoron field obeys the Klein-Gordon equation:
\begin{equation}
\ddot{J} + 3H\dot{J} + V_J = 0 \ , \label{eq: KG}
\end{equation}
where the dot denotes the time derivative, $H \equiv \dot{a}/a$ is the Hubble paremeter and $V_J \equiv dV/ dJ$.
For our convenience, we define dimensionless time and field variables:
\begin{equation}
t_m \equiv m_J t \ , \qquad \theta \equiv J/F_J \ .
\end{equation}
Then, from \eqref{eq: KG} we obtain the equation of motion for $\theta$:
\begin{equation}
\theta'' + 3\dfrac{H}{m_J}\theta' + \sin\theta = 0 \ , 
\end{equation}
where the prime denotes a time derivative with respect to $t_m$.
Regarding the time evolution of $J$, we assume that $J$ is initially frozen on the potential due to a large Hubble friction.
However, since the Hubble parameter decreases in time, at a certain time $t_m = t_{m, \rm osc}$ Hubble friction becomes comparable with a gradient of potential and Majoron starts to oscillate around its minimum, therby behaving as a non-relativistic fluid.

We determine $t_{m,osc}$ by numerics for each of the initial conditions.
In the misalignment mechanism, the relic abundance depends on the initial field value. 
As usual for the quadratic potential, $t_{m,osc} = \mathcal{O}(1)$ corresponds to $m_J \sim 3H$.
On the other hand, near the hilltop of the cosine potential, anharmonic effects can significantly enhance the abundance \cite{Kobayashi:2013nva,Co:2018mho}.
As a result, the observed dark matter density can be achieved even for relatively smaller decay constants, thereby opening up parameter regions with larger couplings that may be accessible to experiments.
For $t_{m,osc}$ with an initial condition close to the top of cosine potential, we solve it by denoting a deviation from $\theta = \pi$ as
\begin{equation}
\delta\theta \equiv \pi -\theta \ .
\end{equation}
For a small deviation $|\delta\theta| \ll 1$, the equation of motion for $\delta\theta$ is approximated as
\begin{equation}
\delta\theta'' + 3\dfrac{H}{m_J}\delta\theta' - \delta\theta \simeq 0 \ . \label{eq: eomdelta}
\end{equation}
Assuming the radiation-dominated era, $H = 1/(2t)$, we can solve this equation exactly.
For finite values of the initial condition, we get
\begin{equation}
\delta\theta = C(-it_m)^{-1/4}J_{1/4}(-i t_m) \ , \label{eq: deltasol}
\end{equation}
where $J_\nu(x)$ is the Bessel function and $C$ is an integration constant:
\begin{equation}
C = 2^{1/4}\Gamma(\tfrac{5}{4})\delta\theta_i
\end{equation}
with an initial angle deviation $\delta\theta_i$.
For large $t_m$, \eqref{eq: deltasol} is approximated as
\begin{equation}
\delta\theta \simeq \dfrac{C}{ \sqrt{2\pi}}\dfrac{e^{t_m}}{t_m^{3/4}} \ . \label{eq: deltatheta}
\end{equation}
For small $\delta\theta_i$, the solution for $t_m$ is given by the negative branch of Lambert W-function:
\begin{equation}
t_m = -\dfrac{3}{4}W_{-1}\left(-\dfrac{4}{3}\left(\dfrac{\sqrt{2\pi}\delta\theta}{C}\right)^{-4/3}\right) \ . \label{eq: A}
\end{equation}
Finally, we can roughly evaluate $t_{m,osc}$ as a time when $\delta\theta$ becomes unity $\delta\theta = 1$.

\begin{figure}[t]
\begin{center}
    \includegraphics[clip, width=0.7\columnwidth]{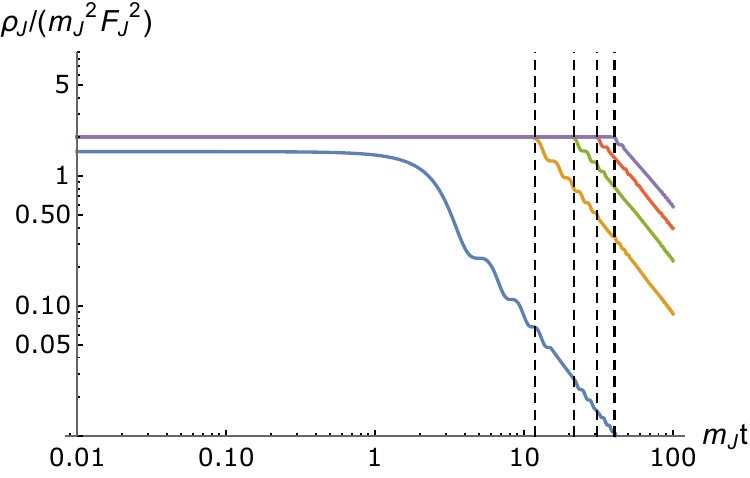}
\caption{
Time evolution of the energy density of Majoron field with $\delta\theta_i = 1$ (blue), $\delta\theta_i = 10^{-4}$ (yellow), $\delta\theta_i = 10^{-8}$ (green), $\delta\theta_i = 10^{-12}$ (red), and $\delta\theta_i = 10^{-16}$ (purple). The dashed lines are corresponding values of $t_{m,osc} = t_m(\delta\theta=1)$ in \eqref{eq: A}.
}
\label{fig: rhoevo}
\end{center}
\end{figure}

We plot the time evolution of the energy density of Majoron field in Figure \ref{fig: rhoevo}.
One can find that the hilltop initial values delay their oscillation times and keep current dark matter abundance even with the steep potential slope (smaller values of $F_J$).
Namely, it enhances the coupling constant in the parameter space of dark matter.
However, the value of $t_{m,osc}$ is an order of 10 even for very tiny values of $\delta\theta_i$.
As has already been discussed in the previous work \cite{Co:2018mho}, we can see that $t_{m,osc}$ is not sensitive to the fine-tuning of $\delta\theta_i$ but it varies only logarithmically.

To obtain the coupling constant of Majoron dark matter to photon, we evaluate the abundance of background Majoron field at present.
The Majoron density parameter $\Omega_J$ is defined as
\begin{equation}
\Omega_J \equiv \dfrac{\rho_J}{3M_p^2H_0^2} \ , \qquad \rho_J \equiv \dfrac{1}{2}\dot{J}^2 + V(J) \ ,
\end{equation}
where $H_0 = 100h \ \text{km} \ \text{s}^{-1} \ \text{Mpc}^{-1}$ is Hubble constant with a dimensionless Hubble parameter $h \simeq 0.7$.
From the above time evolution, its solution is given by a ratio of scale factor between the oscillation time and present time $t = t_0$:
\begin{equation}
\Omega_J \simeq \left(\dfrac{a(t_{\rm osc})}{a(t_0)}\right)^3\dfrac{\rho_J(t_{\rm osc})}{3M_p^2H_0^2} \simeq \left(\dfrac{a(t_{\rm osc})}{a(t_0)}\right)^3\dfrac{m_J^2F_J^2}{3M_p^2H_0^2}\left[ 1 - \cos \theta(t_{osc})\right] \ .
\end{equation}
We assume $a(t_0) = 1$.
For the Majoron mass of our interest, Majoron starts to oscillate during the radiation dominated era.
At this epoch, $a(t_{\rm osc})$ is approximately given by
\begin{equation}
a(t_{\rm osc}) \simeq (2\Omega_{r0}^{1/2}H_0t_{\rm osc})^{1/2} \ ,
\end{equation}
where $\Omega_{r0}h^2 \simeq 2.47\times10^{-5}$ is a radiation density parameter.
Then, we can evaluate the corresponding Majoron-photon coupling constant:
\begin{align}
g_{J\gamma} &= \dfrac{\alpha c_{J\gamma}}{\sqrt{3}\pi M_p}(\Omega_Jh^2)^{-1/2}(\Omega_{r0}h^2)^{3/8}(2t_{m,osc})^{3/4}\left(\dfrac{m_Jh}{H_0}\right)^{1/4}\left[ 1 - \cos \theta(t_{osc})\right]^{1/2} \ .
\end{align}
For the hilltop initial condition, $\cos \theta(t_{osc}) \simeq -1$, it is evaluated as
\begin{align}
g_{J\gamma} &\simeq 4.4\times 10^{-15} \text{GeV}^{-1}\left(\dfrac{n}{8}\right)\left(\dfrac{t_{m,osc}}{10}\right)^{3/4}\left(\dfrac{\Omega_Jh^2}{0.120}\right)^{-1/2}\left(\dfrac{m_J}{10^{-10}\text{eV}}\right)^{1/4} \ .
\end{align}
In terms of $F_J$, we obtain
\begin{equation}
F_J = c_{J\gamma}\dfrac{\alpha}{\pi}g_{J\gamma}^{-1} \simeq 1.0\times10^{14}\text{GeV} \left(\dfrac{t_{m,osc}}{10}\right)^{-3/4}\left(\dfrac{\Omega_Jh^2}{0.120}\right)^{1/2}\left(\dfrac{m_J}{10^{-10}\text{eV}}\right)^{-1/4} \ ,
\end{equation}
which is a well-motivated right-handed neutrino mass scale \cite{Buchmuller:2005eh}.

\section{Constraints on the parameter space from gravitational wave detectors} \label{section: main result}

In this section, we present the methods developed in \cite{Nagano:2019rbw,Nagano:2021kwx} to test photon birefringence induced by dark matter with a resonant cavity experiment, and compare the predicted Majoron-photon coupling with the experimental sensitivity of current and upcoming gravitational wave detectors.
In the presence of Chern-Simons interaction \eqref{eq: CS}, the dispersion relations of circular polarization photons are modified by the background Majoron field
\begin{equation}
\omega_{L/R}^2 = k^2\left(1 \mp g_{J\gamma}\dot{J}/k\right)
\end{equation}
and the phase velocity difference of circularly polarized photons is generated:
\begin{equation}
\delta c(t) \equiv \dfrac{1}{2}(c_R - c_L) \simeq \dfrac{g_{J\gamma}\dot{J}(t)}{2k} \ .
\end{equation}
The time evolution of dark matter field is given by
\begin{equation}
J(t) = \dfrac{\sqrt{2\rho_{\rm DM}}}{m_J}\sin(m_Jt + \delta(t)) \ ,
\end{equation}
where $\rho_{\rm DM} \simeq 0.4 \ \text{GeV cm}^{-3}$ is the local abundance of energy density of dark matter \cite{Iocco:2015xga,Sivertsson:2017rkp}.
Majoron field oscillates with a frequency of Majoron mass $f_J = m_J/(2\pi) \simeq 2.4 \text{Hz}(m_J/10^{-14}\text{eV})$.
The phase factor $\delta(t)$ generically depends on time due to the velocity dispersion of dark matter and loses the coherence of the oscillatory motion of the dark matter field \cite{Nakatsuka:2022gaf}.

\begin{figure}[thpb]
\begin{center}
    \includegraphics[clip, width=1.0\columnwidth]{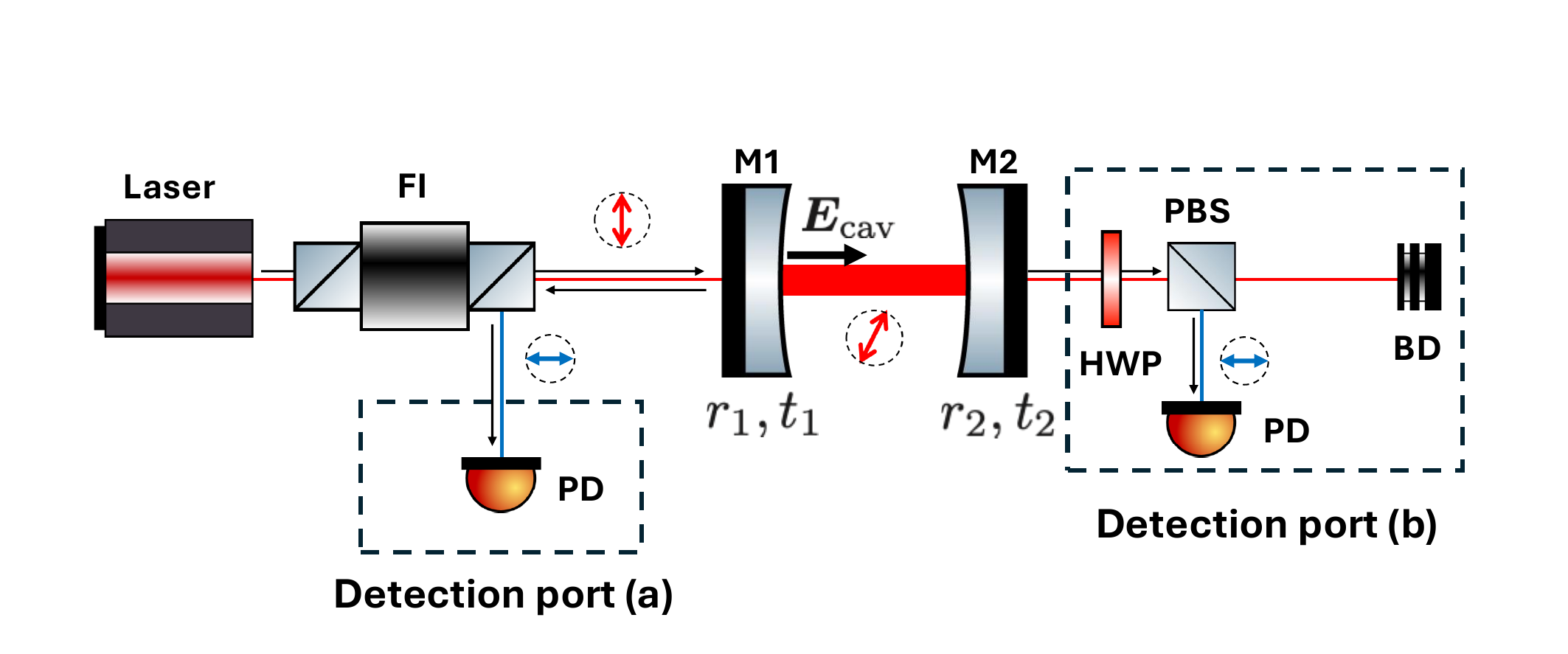}
\caption{
Schematic diagram of a dark matter experiment: FI: Faraday isolator; M1(2): mirror with reflectivity $r_{1(2)}$ and transmittivity $t_{1(2)}$; $\bm{E}_{\rm cav}$: electric field inside the cavity at the reflection mirror M1; HWP: half wave plate; PBS: polarizing beam splitter; BD: beam dump; PD: photodetector. The polarization plane of the incident laser rotates inside the resonant cavity due to the interaction between dark matter and photons. The laser polarization is separated using a polarizing beam splitter and a new polarization state generated by dark matter is measured in the detection port near the cavity reflection mirror (a) or transmission mirror (b).
}
\label{fig: cavity}
\end{center}
\end{figure}

Then, the velocity difference induced by the dark matter leads to a phase difference of left- and right-handed electric fields $\delta\phi$ in the resonant cavity.
We illustrate a schematic experimental setup to detect this signal in Figure \ref{fig: cavity}.
By converting the circular polarization basis into the linear polarization basis, we can find that $\delta\phi$ corresponds to an amplitude of linear polarization.
Assuming the p-polarization is an incident laser light, the electric field inside the resonant cavity at the reflection mirror is approximately given by
\begin{equation}
\bm{E}_{\rm cav} \simeq \dfrac{t_1}{1-r_1r_2}\left[ \bm{E}^p - \delta\phi\bm{E}^s\right] \ ,
\end{equation}
where $r_i(t_i)$ ($i=1,2$) is the reflectivity (transmittivity) of the mirrors, $\bm{E}^{p(s)}$ is an electric vector of p(s)-polarization.
Transforming $\delta c$ into Fourier space $\delta c(t) = \int\tfrac{dm}{2\pi}\tilde{\delta c}(m)e^{imt}$, the induced phase factor in the resonant cavity is decomposed as
\begin{equation}
\delta\phi(t) = \int\dfrac{dm}{2\pi}\tilde{\delta c}(m) H(m)e^{imt} \ .
\end{equation}
Here, $H(m)$ is a response function which transfers the light velocity difference $\delta c$ to the dark matter induced linear polarization amplitude $\delta\phi$.
The mass dependence on the response function differs from which detection ports we install the optics.
The response function at the detection port (a) near the reflection mirror is given by \cite{Nagano:2019rbw}
\begin{equation}
H_{(a)}(m) = i\dfrac{k}{m}\dfrac{4r_1r_2\sin^2(m L_{\rm cav}/2)}{1-r_1r_2e^{-i2m L_{\rm cav}}}(-e^{-im L_{\rm cav}}) \ ,
\end{equation}
where $L_{\rm cav}$ is a cavity length.
On the other hand, the response function at the detection port (b) near the transmission mirror is given by \cite{Nagano:2021kwx}
\begin{equation}
H_{(b)}(m) = H_{(a)}(m) + H_t(m)
\end{equation}
with the contribution from one way translation of later light in the cavity: 
\begin{equation}
H_t(m) = \dfrac{2k}{m}e^{i mL_{\rm cav}/2}\sin\left(\dfrac{mL_{\rm cav}}{2}\right) \ .
\end{equation}

About experimental noise, assuming that the quantum shot noise is the primary noise source, we obtain the one-sided noise spectrum by comparing the signal $\tilde{\delta c}(m)$ with the vacuum fluctuation of electric field \cite{Kimble:2000gu}
\begin{equation}
\sqrt{S_{\rm shot}(m)} = \dfrac{1}{\tfrac{t_1t_i}{1-r_1r_2}|H(m)|\sqrt{\tfrac{2P_0}{k}}} \ ,
\end{equation}
where $P_0$ is an incident laser power and $k$ is a wave number of laser light.
Note that in this expression the vacuum fluctuation of the electric field is normalized as unity and the dimension of the electric field is taken $\rm Hz^{1/2}$.
Then, the signal-to-noise ratio (SNR) is given by
\begin{equation}
\rm SNR = \begin{cases}
\dfrac{\sqrt{T_{\rm obs}}}{2\sqrt{S_{\rm shot}}}\delta c_0 \quad (T_{\rm obs} \lesssim \tau) \\
\dfrac{(\tau T_{\rm obs})^{1/4}}{2\sqrt{S_{\rm shot}}}\delta c_0 \quad (T_{\rm obs} \gtrsim \tau) \ ,
\end{cases}
\end{equation}
where its improvement with a measurement time $T_{\rm obs}$ depends on the magnitude of coherent time scale of dark matter $\tau \equiv 2\pi/(m_Jv^2) \sim 1(m_J/10^{-16}\text{eV})^{-1}\text{yrs}$.
By setting SNR to unity, we obtain the corresponding sensitivity for the Majoron dark matter-photon coupling constant:
\begin{equation}
g_{J\gamma} \simeq 1.9\times10^{12}\text{GeV}^{-1}\left(\dfrac{\lambda}{1064\text{nm}}\right)^{-1}\dfrac{\sqrt{2\rho_{\rm DM}}}{m_J}
\times\begin{cases}
\sqrt{\dfrac{S_{\rm shot}}{T_{\rm obs}}} \quad (T_{\rm obs} \lesssim \tau) \\
\dfrac{\sqrt{S_{\rm shot}}}{(\tau T_{\rm obs})^{1/4}} \quad (T_{\rm obs} \gtrsim \tau) \ ,
\end{cases}
\end{equation}
where $\lambda = 2\pi/k$ is the wavelength of the laser light.

\begin{figure}[thpb]
\begin{center}
    \includegraphics[clip, width=0.7\columnwidth]{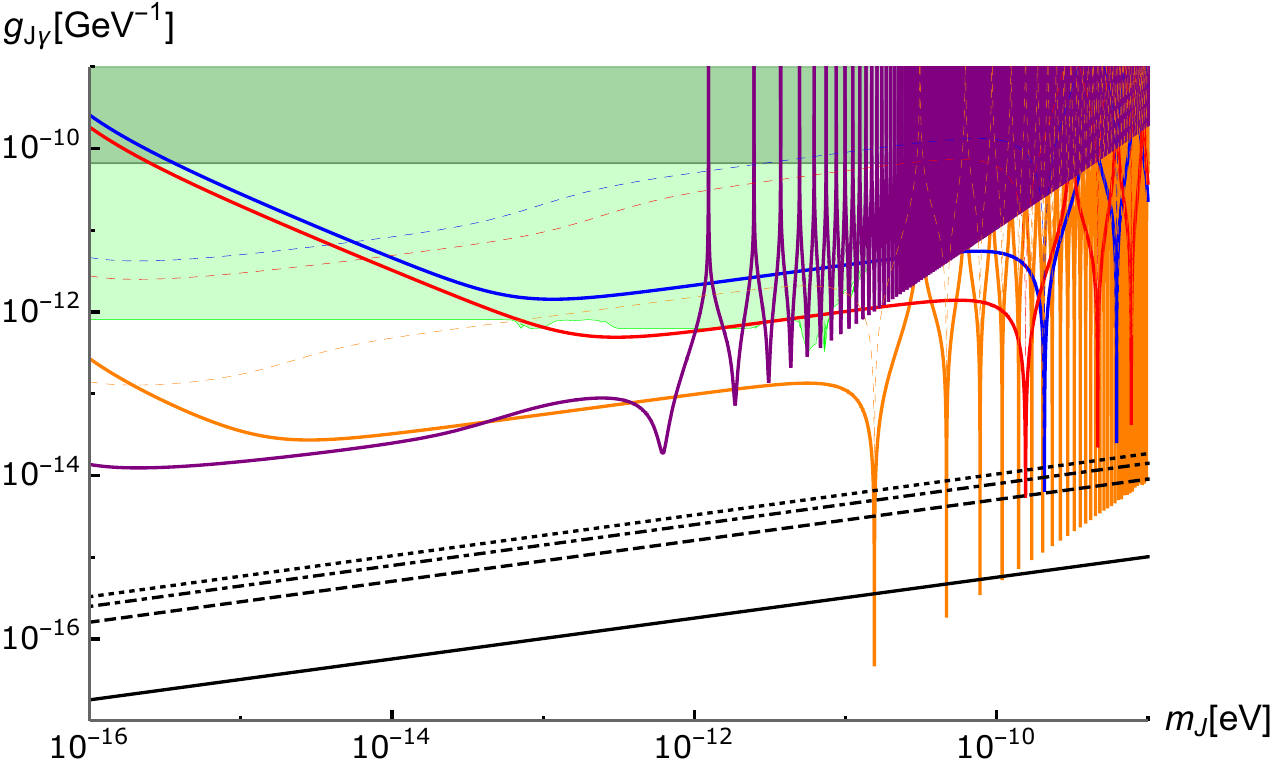}
\caption{
Sensitivity curves of several gravitational wave detectors for the Majoron-photon coupling constant with respect to Majoron dark matter mass: KAGRA (blue), aLIGO (red), DECIGO (purple) and Cosmic Explorer (yellow) with $n=8$, $T_{\rm obs} = 1 \text{yr}$ and with several parameter sets in Table \ref{tab:ITFparameters}. 
The solid curves are the use of detection port (a), while the dashed curves are the detection port (b).
The green contours show the current exclusion limit from axion-photon conversion experiment and astrophysical observation: CAST \cite{CAST:2017uph} (dark green) and NGC1275 by Chandra \cite{Reynolds:2019uqt} (light green).
The black lines show the parameter region of Majoron dark matter with a cosine potential with $\theta_J=\mathcal{O}(1)$ (solid) and that with hilltop initial conditions: $\delta\theta_i = 10^{-4}$ (dashed), $\delta\theta_i = 10^{-8}$ (dotdashed), $\delta\theta_i = 10^{-12}$ (dotted).
}
\label{fig: mainresult}
\end{center}
\end{figure}

In Figure \ref{fig: mainresult}, we plot the parameter space of Majoron dark matter mass and Majoron-photon coupling constant, and compare the sensitivity curves of gravitational wave detectors, with the use of model parameters in Table \ref{tab:ITFparameters}.
In this plot, we assume 1-year observation and Majoron is a dominant component of dark matter.
For detection port (b), the sensitivity is better on a lower dark matter mass due to the value of one way transformation of light travel \cite{Nagano:2021kwx}.
On the other hand, the detection port (a) is better for the higher dark matter mass region with sharp peaks corresponding to the free spectral range of detectors: $m_J = (2N-1)\pi/L_{\rm cav} \ (N \in \mathbf{N})$.
The black solid line is a parameter region of Majoron dark matter with a typical initial angle of cosine potential: $\theta_i =\mathcal{O}(1)$.
We can see that the sensitivity level of future gravitational wave detectors such as Cosmic Explorer (CE)-like experiment could probe with several narrow bands corresponding to a free spectral range, while it is hard to cover a wide dark matter mass region.
On the other hand, for hilltop initial conditions broader dark matter mass range becomes testable, even with current sensitivity levels such as Advanced LIGO (aLIGO) and KAGRA.

Such a hilltop initial condition could be realized if the primordial perturbation is sufficiently small comparing with the initial angle deviation from the top of potential\footnote{Regarding the particular choice of an initial angle very close to the hilltop, several studies have explored the possibility of realizing such a hilltop misalignment mechanism in the early Universe \cite{Co:2018mho,Daido:2017wwb,Takahashi:2019pqf}.}.
The magnitude of inflationary fluctuation is given by $H_i/(2\pi)$, where $H_i$ is an inflationary Hubble scale.
For instance, given that $H_i$ is around $\mathcal{O}(\text{GeV})$ scale, we obtain
\begin{equation}
\delta\theta_i \gg \dfrac{H_i}{2\pi F_J} \simeq 1.6 \times 10^{-15}\left(\dfrac{H_i}{\text{GeV}}\right)\left(\dfrac{10^{14}\text{GeV}}{F_J}\right) \ .
\end{equation}
For an extreme initial value $\delta\theta_i \ll 1$, however, isocurvature perturbation is significantly enhanced due to the anharmonic effect of the potential \cite{Kobayashi:2013nva}.
To ensure that the isocurvature perturbation is subdominant to the adiabatic curvature perturbation, we estimate
\begin{equation}
S \sim \dfrac{H_i}{2\pi F_J\delta\theta_i} \ll 10^{-5} \qquad \longleftrightarrow \qquad H_i \ll 6.3 \ \text{GeV}\left(\dfrac{F_J}{10^{14}\text{GeV}}\right)\left(\dfrac{\delta\theta_i}{10^{-9}}\right) \ .
\end{equation}
Therefore, an extremely low inflationary Hubble scale is required to satisfy the observational constraints\footnote{This isocurvature problem can be avoided if the field is dynamically driven and stabilized at the potential minimum during inflation \cite{Co:2018mho}.}.
On the other hand, if the inflation scale is high as $H_i =O(10^{13})$ GeV for $\theta_i = \mathcal{O}(1)$, we also encounter a large isocuvature problem. 
However, this problem can be solved if the field value is initially taken a large value such as Planck scale \cite{Linde:1991km} and $F_J \gtrsim O(10^{14})$ GeV is required \cite{Kawasaki:2018qwp}. 
Thus, by relying on this mechanism, the parameter region discussed in the present paper would not have the isocuvature problem even if the inflation scale is high.

\section{Discussion and conclusion} \label{section: conclusion}

We have developed an anomalous Majoron model in which the Majoron behaves as dark matter, and explored the parameter space of the Majoron-photon coupling and the dark matter mass with the optical cavity experiment based on the gravitational wave detectors.
We have found that the sensitivity level of future ground-based laser interferometers potentially probe a parameter region of the Majoron dark matter with mass around $10^{-10}$ \text{eV}.
\begin{table*}[htb]
\centering
 \caption{Parameters of considered gravitational wave detectors: the cavity length $L_{\rm cav}$, the input beam power to the cavity $P_0$, the laser wavelength $\lambda$ and the mirror transmittivities $(t_1, t_2)$.
 }
 \label{tab:ITFparameters}
 \begin{tabular}{c|c|c|c|c}
  Similar detector      &       $L_\mathrm{cav}$     [m] &   $P_0$ [W]       &       $\lambda$ [$\times 10^{-9}$ m]    &       $(t_1^2, t_2^2)$ [ppm]  
  \\ \hline
  KAGRA~\cite{Somiya:2011np}        &       $3\times10^3$   &       335    &       1064    &       ($4\times10^3$, 7)\\ \hline
  aLIGO~\cite{LIGOScientific:2014pky}        &       $4\times10^3$   &       2600    &       1064    &       ($1.4\times10^4$, 5)\\ \hline
  CE~\cite{LIGOScientific:2016wof}       &       $4\times10^4$   &       600    &       1550    &       ($1.2\times10^3$, 5)  \\ \hline 
  DECIGO~\cite{Kawamura:2008zz}    &       $10^6$  &       5       &       515     &       ($3.1 \times 10^5$, $3.1 \times 10^5$)       
 \end{tabular}
\end{table*}
If the initial field value is close to the hilltop of the Majoron potential, we may have a chance to test a wider dark matter mass range even for the current sensitivity levels such as aLIGO or KAGRA.
To probe it, we need to install additional optics to extract photon birefringence in the detector site.
We have shown that the response at a detection port located near the reflection mirror provides better sensitivity in the higher-mass region compared to that near the transmission mirror.
However, incorporating optics on the side of the reflection mirror is technically challenging because gravitational wave signals are detected on this side.
As a future direction, it will be necessary to perform numerical simulations that include the effects of noises which are expected to arise once we install such optics for dark matter detection.
Moreover, in this work we have considered only one arm of the laser interferometer, and a full evaluation of the signal response including both arms has not yet been carried out.
Therefore, further dedicated studies are required to realize this detection scheme.

In this paper, we had a drawback in the previous anomalous Majoron model \cite{Liang:2024vnd}, where there are two Higgs doublets $H_1$ and $H_2$ and the masses of the hierarchy between $\mathcal{O}(10^{14})$ GeV and electroweak scale $\mathcal{O}(10^2-10^3)$ GeV were imposed.
However, as long as we consider a mass mixing of Higgs fields, we would need serious fine tuning among their mass-matrix elements.
This is a result of the choice of the coupling $H_2^\dagger H_1 \Phi$ with $F_J \sim 10^{14}$ GeV.
But, this problem can be potentially solved if we use a higher dimensional operator such as $H_2^\dagger H_1 (\Phi^n/ M^{n-2}_{PL})$. 
Considering the mass matrix of two Higgs doublets, the current model is advantageous assuming that both of diagonal terms $M_{1,2}^2$ are positive.
If both of $M_{1,2}^2$ are much greater than the mixing mass $\delta^2$, the electroweak symmetry is never broken down.
And if one of $M_{1,2}^2$ is much greater than $\delta^2$, the mixing of the two Higgs doublets is very small and hence all masses of quarks or leptons become too small. Thus, all of masses for $M_{1,2}^2$ and $\delta^2$ are most likely at the same magnitudes. Namely, the electroweak breaking scale is given by the magnitude of the $\delta^2$. 
On the other hand, the magnitude of the $\delta^2$ is determined by the $U(1)_N$ symmetry breaking scale $F_J$ and the charges of the Higgs doublets $H_2^\dagger H_1$ (which determines $n$). 
In other words, the electroweak breaking scale is determined by the $U(1)_N$ charge of the $H_2^\dagger H_1$ and the $U(1)_N$ breaking scale $F_J$.
Furthermore, it is even more interesting if we may have a chance to find the second Higgs boson in future experiments.

\section*{Acknowledgement}

We thank Yuta Michimura for fruitful discussions.
This work of I.O. is supported by JSPS KAKENHI Grant Nos. 19K14702 and by KEK support program for young resercher No. B0E510521. This work of  T. T. Y. is supported by the JSPS KAKENHI Grants No. 24H02244 and the World Premier International Research Center Initiative (WPI), MEXT, Japan (Kavli IPMU).

\bibliographystyle{utphys}
\bibliography{Refrences.bib}

\end{document}